\input epsf
\input harvmac

\lref\bls{L. Bombelli, R.K. Koul, J. Lee and R.D. Sorkin, {\it
Quantum source of information for black holes}, Phys.\ Rev.\ {\bf D34}
(1986) 373.}

\lref\gito{G.W. Gibbons and P.K. Townsend, {\it Black Holes and
Calogero Models}, Phys.\ Lett.\ {\bf B454} (1999) 187,
hep-th/9812034.}

\lref\triv{S. P. Trivedi,  {\it Semiclassical extremal black holes},
Phys.\ Rev.\ {\bf D47} (1993) 4233, hep-th/9211011.}

\lref\holzhey{C. Holzhey, Ph.D. Thesis, Princeton University, 1993.}

\lref\larsen{F. Larsen and F. Wilczek, {\it Renormalization of Black
Hole Entropy and of the Gravitational Coupling Constant},
Nucl.\ Phys.\ {\bf B458} (1996) 249, hep-th/9506066.}

\lref\pbm{A.P. Prudnikov, Yu.A. Brychkov and O.I. Marichev, {\it
Integrals and Series:  Volume 2:  Special Functions}, (Gordon and
Breach, 1986).}

\lref\suss{L. Susskind and J. Uglum, {\it Black Hole Entropy in
Canonical Quantum Gravity and Superstring Theory}, Phys.\ Rev.\ 
{\bf D50} (1994) 2700, hep-th/9401070.}

\lref\kabat{D. Kabat, {\it Black Hole Entropy and Entropy of
Entanglement}, Nucl.\ Phys.\ {\bf B453} (1995) 281, hep-th/9503016.}

\lref\mark{M. Srednicki, {\it Entropy and Area}, Phys.\ Rev.\ Lett.\ 
{\bf 71} (1993) 666, hep-th/9303048.}

\lref\sttr{A. Stominger and S.P. Trivedi, {\it Information
Consumption by Reissner-Nordstrom Black Holes}, Phys.\  Rev.\ 
{\bf D48} (1993) 5778, hep-th/9302080.}

\lref\kim{W.T. Kim, {\it AdS$_{2}$ and quantum stability in the
CGHS model}, Phys.\  Rev.\  {\bf D59} (1999) 047503, hep-th/9810055.}

\lref\carter{B. Carter, {\it Black Holes}, edited by C. de Witt and
B.S. de Witt, (Gordon and Breach, New York, 1973).}

\lref\lemos{J.P.S. Lemos, {\it Thermodynamics of the Two-dimensional
Black Hole in the Teitelboim-Jackiw Theory}, Phys.\ Rev.\ {\bf D54}
(1996) 6206, gr-qc/9608016.}

\lref\mann{R. B. Mann, {\it Lower Dimensional Black Holes},
General Relativity and Gravitation {\bf 24} (1992) 433.}

\lref\mannste{R. B. Mann and T.G. Steele, {\it Thermodynamics and
quantum aspects of black holes in (1+1) dimensions}, Class.\ Quant.\ 
Grav.\ {\bf 9} (1998) 475.}

\lref\cw{C. Callan and F. Wilczek, {\it On geometric entropy}, Phys.\ 
Lett.\ {\bf B333} (1994) 55, hep-th/9401072.}

\lref\btz{M. Ba\~nados, C. Teitelboim and J. Zanelli, {\it Black Hole
in Three-Dimensional Spacetime}, Phys.\ Rev.\ Lett.\ {\bf 69} (1992)
1849.}

\lref\fks{S. Ferrara, R. Kallosh and A. Strominger, {\it N = 2
Extremal Black Holes}, Phys.\ Rev.\ {\bf D52} (1995) 5412,
hep-th/9508072.}

\lref\dhj{E. D'Hoker and R. Jackiw, {\it Classical and quantal
Liouville field theory}, Phys.\ Rev.\ {\bf D26} (1982) 3517.}

\lref\dhfj{E. D'Hoker, D.Z. Freedman and R. Jackiw, {\it
SO(2,1)-invariant quantization of the Liouville theory}, Phys.\ Rev.\ 
{\bf D28} (1983) 2583.}

\lref\fmmr{D.Z. Freedman, S. D. Mathur, A. Matusis and L. Rastelli,
{\it Correlation functions in the $CFT_d/AdS_{d+1}$ correspondence},
Nucl.\ Phys.\ {\bf B546} (1999) 96,
hep-th/9804058.}

\lref\bdhm{T. Banks, M. R. Douglas, G. T. Horowitz and E. Martinec,
{\it AdS Dymanics from Conformal Field Theory}, hep-th/9808016.}

\lref\fiola{T.M. Fiola, J. Preskill, A. Strominger and S.P. Trivedi,
{\it Black hole thermodynamics and information loss in two
dimensions}, Phys.\ Rev.\ {\bf D50} (1994) 3987, hep-th/9403137.}

\lref\frag{J. Maldacena, J. Michelson and A. Strominger, {\it
Anti-de Sitter Fragmentation}, JHEP 02 (1999) 011,
hep-th/9812073.}

\lref\ffgk{V. Frolov, D. Fursaev, J. Gegenberg and G. Kunstatter,
{\it Thermodynamics and Statistical Mechanics of Induced Liouville
Gravity},
Phys.\ Rev.\ {\bf D60} (1999) 024016,
hep-th/9901087.}

\lref\gkp{S.S. Gubser, I.R. Klebanov and A.M. Polyakov, {\it Gauge
theory correlators from noncritical string theory}, Phys.\ Lett.\ 
{\bf B428} (1998) 105, hep-th/9802109.}

\lref\burgess{C.P. Burgess and C.A. L\"utken, {\it Propagators and
effective potentials in anti-de Sitter space}, Phys.\ Lett.\ 
{\bf B153} (1985) 137.}

\lref\davies{P.C.W. Davies and S.A. Fulling, {\it Quantum vacuum
energy in two dimensional space-time}, Proc.\ R. Soc.\ Lond.\ 
{\bf A354} (1977) 59.}

\lref\barf{W.A. Bardeen and D.Z. Freedman, {\it On the energy crisis
in anti-de Sitter supersymmetry}, Nucl.\ Phys.\ {\bf B253} (1985) 635.}

\lref\gr{I.S. Gradshteyn and I.M Ryzhik, {\it Tables of Integrals,
Series, and Products}, (Academic Press, San Diego, 1994).}

\lref\full{S. A. Fulling, {\it Aspects of Quantum Field Theory in
Curved Space-Time}, (Cambridge University Press, 1989).}

\lref\bd{N.D. Birrell and P.C.W. Davies, {\it Quantum Fields in
Curved Space}, (Cambridge University Press, 1982).}

\lref\albo{V. Balasubramanian, P. Kraus and A. Lawrence, {\it Bulk
vs. Boundary Dynamics in Anti-de Sitter Spacetimes}, Phys.\ Rev.\ 
{\bf D59} (1999) 046003, hep-th/9805171.}

\lref\albt{V. Balasubramanian, P. Kraus, A. Lawrence and S. Trivedi,
{\it Holographic Probes of Anti-de Sitter Spacetimes},
Phys.\ Rev.\ {\bf D59} (1999) 104021,
hep-th/9808017.}

\lref\witten{E. Witten, {\it Anti De Sitter Space and Holography},
Adv.\ Theor.\ Math.\ Phys.\ {\bf 2} (1998) 253, hep-th/9802150.}

\lref\ulf{U.H. Danielsson, E. Keski-Vakkuri and M. Kruczenski, {\it
Vacua, Propagators, and Holographic Probes in AdS/CFT},
JHEP 01 (1999) 002,
hep-th/9812007.}

\lref\juan{J. Maldacena, {\it The Large N Limit of Superconformal
field theories and supergravity}, Adv.\ Theor.\ Math.\ Phys.\ {\bf 2}
(1998) 231, hep-th/9711200.}

\lref\andy{A. Strominger {\it $AdS_2$ Quantum Gravity and String
Theory}, JHEP 01 (1999) 007, hep-th/9809027.}

\lref\toshio{T. Nakatsu and N. Yokoi, {\it Comments on Hamiltonian
Formalism Of $AdS/CFT$ Correspondence}, Mod.\ Phys.\ Lett.\ {\bf A14}
(1999) 147, hep-th/9812047.}

\lref\pkt{P.K. Townsend, {\it The M(atrix) model/$adS_2$
correspondence}, Proceedings of the 3rd Puri workshop on Quantum
Field Theory, hep-th/9903043.}

\lref\cadoni{M. Cadoni and S. Mignemi, {\it Asymptotic symmetries of
AdS$_2$ and conformal group in d = 1}, hep-th/9902040.}

\def\ad{$AdS_2$}
\def\adst{$AdS_2\times S^2$}
\def\p{\partial}
\def\slr{{SL(2,{\bf R})}}

\Title{\vbox{\baselineskip12pt
	\hbox{HUTP-99/A014}
	\hbox{hep-th/9904143}
}}{Vacuum States for \ad\ Black Holes}

\centerline{
	Marcus Spradlin\foot{{\tt spradlin@feynman.harvard.edu}}
	and Andrew Strominger\foot{{\tt andy@planck.harvard.edu}}
}

\bigskip
\centerline{Department of Physics}
\centerline{Harvard University}
\centerline{Cambridge, MA 02138}

\vskip .3in
\centerline{\bf Abstract}
An \ad\ black hole spacetime is an \ad\ spacetime together with a
preferred choice of time.  The Boulware, Hartle-Hawking and
$\slr$ invariant vacua are constructed, together with their
Green functions and stress tensors, for both massive and massless
scalars in an \ad\ black hole.  The classical Bekenstein-Hawking
entropy is found to be independent of the temperature, but at
one loop a non-zero entanglement entropy arises.  This represents a
logarithmic violation of finite-temperature decoupling for \ad\ 
black holes which arise in the near-horizon limit of an
asymptotically flat black hole.  Correlation functions of the $\slr$
invariant
boundary quantum mechanics are computed as functions of the choice of
\ad\ vacuum.

\smallskip

\Date{April 1999}

\newsec{Introduction}

Two-dimensional anti-deSitter space (\ad) has arisen in at least
three distinct but related contexts within string/black hole physics.
The first is as the near-horizon geometry (together with an $S^2$
factor) of the extremal Reissner-Nordstrom solution \carter.
\ad\ is a stable attractor solution of the equations which govern how
the geometry changes as the horizon is approached \fks, and as such
is expected to play a central role in the physics of black holes.
\ad\ made a second appearance in studies of two-dimensional quantum
gravity, where it provides an $\slr$ invariant ground state for
Liouville gravity \refs{\dhj, \dhfj}, and a rich arena for the study
of two-dimensional black holes \refs{\mann \triv \sttr \lemos-\ffgk}.
Most recently it is the black sheep in the family of $AdS/CFT$
dualities \juan, having so far resisted a fully satisfactory
realization of the duality \refs{\andy \frag \toshio  \cadoni \gito -
\pkt}.  One hopes that this can be remedied and that in the process a
clearer relation between the different aspects of \ad\ physics will
emerge.

In this paper we investigate properties of both massive and massless
quantum field theory on an \ad\ background.  In section 2 we review
the appearance of \ad\ in near-horizon black hole geometries.  This
motivates the definition of an \ad\ black hole as an \ad\ spacetime
together with a preferred choice of time.  In section 3 it is shown
in the quantum theory that the choice of time affects the vacuum
state.
We discuss
the Hartle-Hawking, Boulware and $\slr$ invariant \ad\ black hole
vacua and the Hawking temperature
measured by various families of observers.  It is shown that the
vacua defined with respect to Poincar\'e or global time are
equivalent to one another and to the Hartle-Hawking-vacuum.  The
Boulware vacuum,
which is associated to the preferred choice of time, is not in
general equivalent.  Section 4 concerns the
entropy of an \ad\ black hole.  The classical Bekenstein-Hawking
entropy is temperature-independent.  At one loop there is an
entanglement entropy which depends logarithmically on the Hawking
temperature.  This represents a violation of low-energy decoupling
between the asymptotically flat and near-horizon regions of the
black hole at finite temperature.  In section 5 we analyze processes
in which the temperature is changed by sending matter into the black
hole.  In section 6 we turn to massive fields, and give explicit
expressions for the Green functions in the Boulware and
Hartle-Hawking vacua.  The stress-energy expectation values in these
vacua are computed in section 7.  In section 8, motivated by the
$AdS/CFT$ duality, we compute correlation functions of the $\slr$
invariant boundary quantum mechanics in the various \ad\ vacua.

\newsec{\ad\ Black Holes in the Near-Horizon Limit}

In three dimensions, all negative curvature spaces are locally
equivalent to $AdS_3$.  Because of this, for many years it was
believed that black holes did not exist for pure gravity in three
dimensions.  However, BTZ showed that black holes do exist which
differ from $AdS_3$ only by global identifications \btz.  The local
geometry at the black hole horizon is the same as everywhere else,
but it is globally characterized as the surface from behind which
nothing can communicate with infinity.  This differs from higher
dimensional examples in which the geometry has special features at
the horizon.

In two dimensions, all negative curvature spaces are locally \ad.  We
will argue that, much as in three dimensions, \ad\ black holes
nevertheless exist in pure gravity (without dilatons).  Similar
discussions have appeared in \refs{\mann, \kim}.  One way to
describe this is that an \ad\ black hole is \ad\ together with a
choice of (Killing) time $t$ at infinity for which the full region
$-\infty < t < \infty$ does not cover all of the boundary of \ad.
The black hole horizon is then the surface from behind which nothing
can escape to the region $-\infty < t < \infty$.  We will see that
the black holes so defined have characteristic
thermodynamic properties.

\ad\ black holes naturally arise in the near-horizon limits of
Reissner-Nordstrom black holes.  Following the discussion of \frag,
the full magnetically-charged solution is
\eqn\reno{\eqalign{
ds^2 &=- {(r - r_+)(r - r_-) \over r^2} dt^2 +
{r^2 \over (r - r_+)(r - r_-)} dr^2 + r^2 d\Omega_2^2,\cr
F &= Q \epsilon_2,
}}
where $\epsilon_2$ is the volume
element on the unit $S^2$.  The locations of the inner and outer
horizons are related to the Hawking temperature $T_H$
and charge via
\eqn\etl{\eqalign{
Q^2 &= {r_+ r_- \over L_p^2},\cr
T_H &= {r_+ - r_- \over 4 \pi r_+^2},
}}
where $L_p$ is the Planck length.

We now consider, as in \frag, the near-horizon limit
\eqn\nhr{
L_p \to 0,
}
with
\eqn\fgh{
U = {r - r_+ \over L_p^2},~~~~~~~ Q,~T_H~~  {\rm fixed}.
}
The metric then reduces to
\eqn\dfrg{
{ds^2 \over Q^2 L_p^2} = - { U (U + 4 \pi Q^2 T_H) \over Q^4} dt^2 +
{1 \over U (U + 4 \pi Q^2 T_H)} dU^2 +d\Omega_2^2.
}
We note that both the the ADM energy $2M = r^+ + r^-$ and the entropy
$S_{BH} = {\pi r_+^2 \over L_p^2}$ go to $T_H$-independent constants
($M = Q$ and $S_{BH} = \pi Q^2$) in this limit.

The $T_H$ dependence of the metric can be eliminated by a coordinate
transformation.  Defining
\eqn\ctf{
t^\prime \pm {Q^2 \over U^\prime} = \tanh \left[ \pi T_H \left(t \pm
{1 \over 4 \pi T_H} \ln {U \over U+4 \pi Q^2 T_H}\right) \right],
}
the metric reduces to
\eqn\dfr{
{ds^2 \over Q^2 L_p^2} = - {U^{\prime 2} \over Q^4} dt^{\prime 2} +
{1 \over U^{\prime 2}} dU^{\prime 2} + d\Omega_2^2.
}
In terms of
\eqn\rgo{
\tau \pm \sigma \pm {\pi \over 2} = 2 \tan^{-1} \left(t^\prime \pm
{Q^2 \over U'}\right),
}
the metric becomes
\eqn\dffr{
{ds^2 \over Q^2 L_p^2} = {-d\tau^2 + d\sigma^2 \over \cos^2 \sigma}
+ d\Omega_2^2.
}
This is known as the Robinson-Bertotti geometry on \adst.  As
illustrated in figures 1a and 1b for the extremal and near extremal
cases, the \adst\ region of the full Reissner-Nordstrom geometry is a
ribbon which zigzags its way up through the infinite chain of
universes.

\smallskip
\centerline{\epsfxsize=0.61\hsize\epsfbox{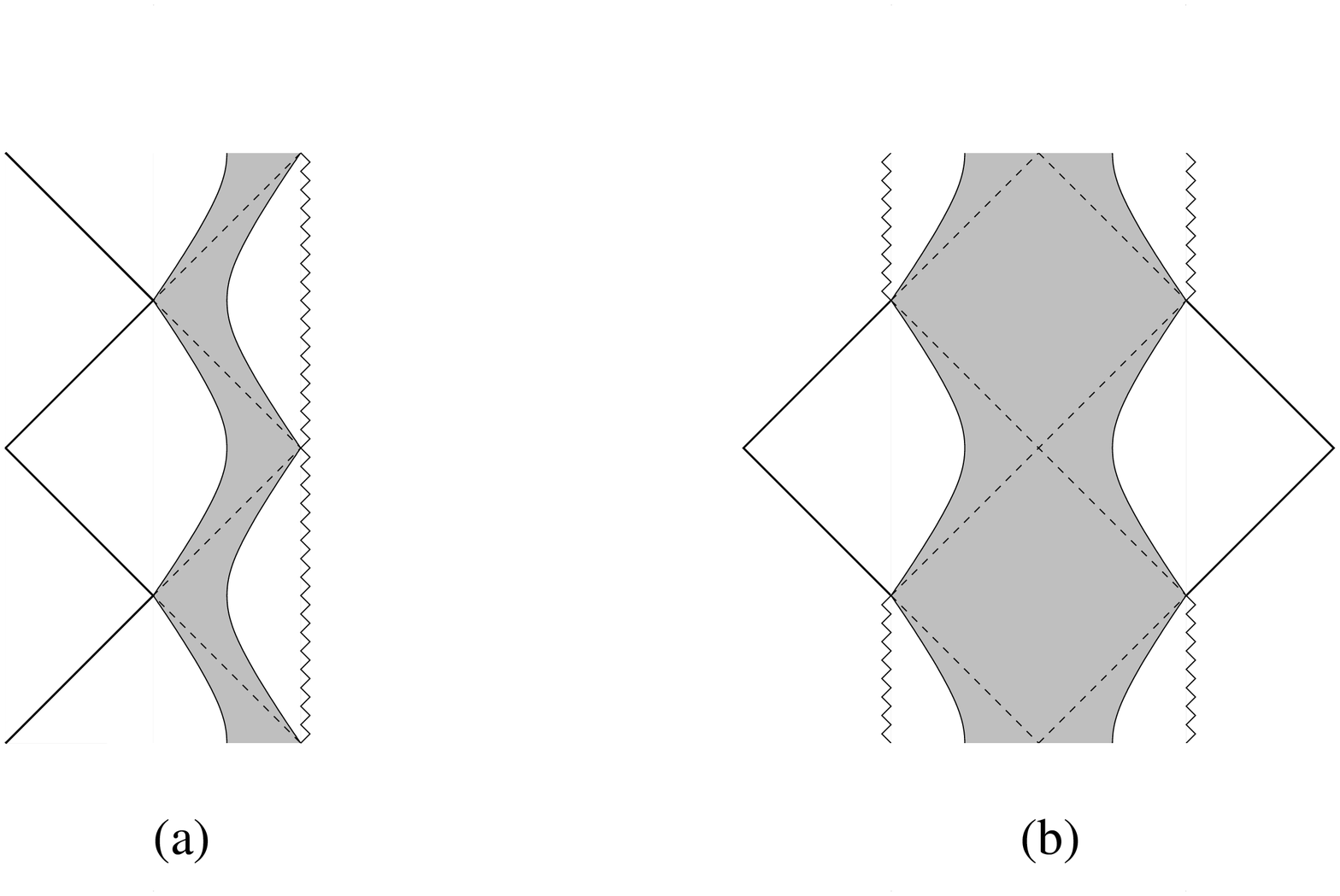}}
\bigskip
\centerline{\vbox{\noindent {\bf Fig.\ 1.}
(a) Penrose diagram corresponding to the extremal
Reissner-Nordstrom black hole.  The dashed line is the black hole
horizon, and the shaded strip is the near-horizon \ad\ region.
(b) Near-extremal Reissner-Nordstrom black hole and corresponding
near-horizon \ad\ region.
}}
\vskip .7cm

Since $T_H$ can be eliminated by a coordinate transformation, the
classical near horizon theory is independent of $T_H$.  We shall see
in the next section
that this is not the case in the quantum theory, because the
definition of a vacuum state in general depends on a choice of time,
or equivalently a preferred family of observers.
An \ad\ spacetime which arises as the near horizon geometry of
Reissner-Nordstrom is indeed endowed with a preferred choice of time
``$t$'', namely, the one associated to the Killing vector which
generates unit time translations in the asymptotically flat spatial
infinity of the Reissner-Nordstrom geometry.  As is evident from
figure 2, as this preferred time coordinate $t$ runs over the full
range $-\infty < t < +\infty$, only part of of the timelike
boundaries of \ad\ is covered.  We shall refer to this boundary
region as spatial infinity.  The future black hole horizon can then
be defined as the boundary of the region from which nothing can
escape to spatial infinity.  The past horizon is then the boundary of
the region which cannot be accessed from spatial infinity.  These
horizons coincide with the Killing horizon of the preferred Killing
vector.

In the extremal case $T_H = 0$ depicted in figure 2a, the exterior of
the black hole is a wedge, the corner of which extend to the far
boundary of \ad.  For $T_H \neq 0$ (figure 2b), the exterior of the
black hole is still a wedge, but it extends only halfway across \ad.

\vskip .6cm
\centerline{\epsfxsize=0.37\hsize\epsfbox{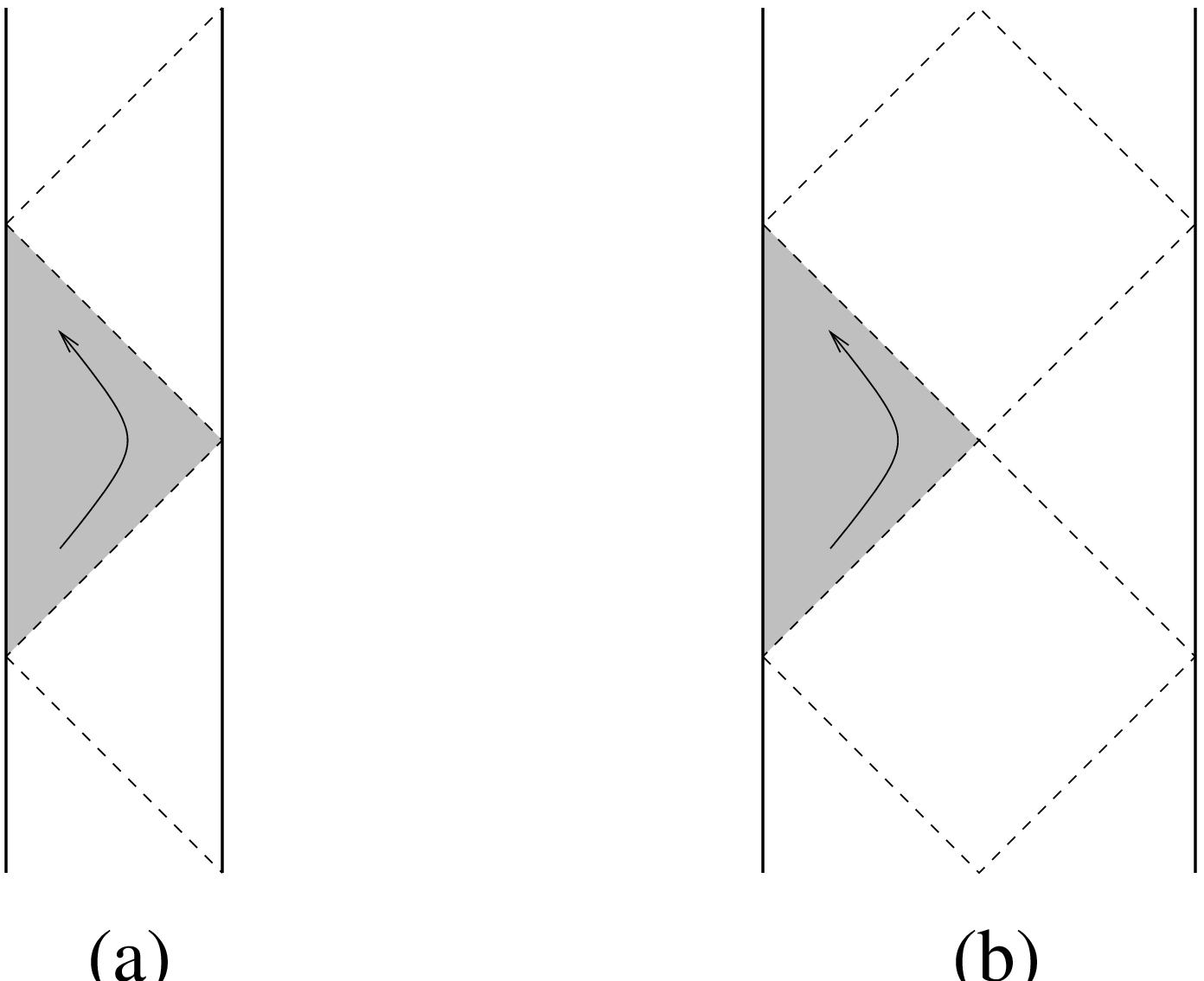}}
\bigskip
\centerline{\vbox{\noindent {\bf Fig.\ 2.}
Penrose diagram for \ad.  The dashed lines are the
horizons inherited from the embedding in extremal (a) or
near-extremal (b) Reissner-Nordstrom (compare with figure 1).
The arrows indicate the flow of asymptotic time ``$t$''.
}}
\vskip .7cm

\newsec{\ad\ Black Hole Thermodynamics}

In this section we discuss the thermal properties of \ad\ black
holes.  We consider mainly the case of a free massless scalar field,
deferring the massive case to section 6.

\subsec{The quantum state}

In order to define a vacuum state we need a metric with a timelike
Killing vector.  The vacuum is then defined as the state annihilated
by positive frequency modes of the field operator.  Observers at a
fixed spatial coordinate $x$, in a coordinate system in which the
metric is time-independent, then detect no particles.

For \ad\ there are inequivalent choices of time coordinates or
equivalently conformal gauge coordinates.  For one such coordinate
choice the metric takes the form
\eqn\bgf{
{ds^2 \over Q^2 L_p^2} = {-d\tau^2 + d\sigma^2 \over \cos^2 \sigma}.
}
The coordinates $(\tau,\sigma)$ are referred to as global coordinates
because they cover all of (the universal cover of) \ad\ for $-{\pi
\over 2} \le \sigma \le {\pi \over 2}$ and $-\infty < \tau < \infty$.
Spatial infinity is at $\sigma = \pm {\pi \over 2}$, and the horizons
are at $\tau \pm \sigma = 0$.  The corresponding vacuum
$|0_{\rm Global}\rangle$, annihilated by modes which are positive
frequency with respect to $\tau$,  is the familiar $\slr$ invariant
vacuum for a free scalar field on the strip.  We shall see shortly
that this is equivalent to the Hartle-Hawking black hole vacuum as
well as the Poincar\'e vacuum.

A second coordinate system is the ``Schwarzschild'' coordinates,
which uses the time $t$ appearing in \dfrg.  $t$ coincides with the
time coordinate inherited from the decoupled asymptotically flat
region and, as discussed above, defines the black hole horizon.
\dfrg\ can be transformed to conformal gauge by
\eqn\ctsf{
x = {1 \over 4 \pi T_H} \ln {U \over U + 4 \pi Q^2 T_H},
}
in which
\eqn\bgfe{
{ds^2 \over Q^2 L_p^2} = \left[{2 \pi T_H \over \sinh (2 \pi T_H x)}
\right]^2 (-dt^2 + dx^2).
}
Since the coordinate transformation \ctsf\ involves only the spatial
coordinate and does not change the choice of time, it does not affect
the associated vacuum $|0_{\rm Schwarzschild} \rangle$.

The Schwarzschild coordinates $(t,x)$ and global coordinates $(\tau,
\sigma)$ are related by the coordinate transformation
\eqn\trans{
\tan \half(\tau \pm \sigma) = \mp e^{\mp 2 \pi T_H (t \pm x)}.
}

A natural family of observers are those moving along worldlines of
fixed $U$.  This corresponds to trajectories which remain a fixed
distance from the black hole horizon.  Since the proper time along
such worldlines equals Schwarzschild time (up to a constant), such
observers will not detect any particles in the state
$|0_{\rm Schwarzschild} \rangle$.  The vacuum with this property is
known as the Boulware vacuum.  Hence we conclude that
\eqn\rop{
|0_{\rm Schwarzschild} \rangle = |0_{\rm Boulware} \rangle.
}
We will see in section 7 that this vacuum has the property that the
expectation value of the stress tensor diverges on the horizon.

Since Schwarzschild and global time do not agree, constant-$U$
observers will detect particles in the global vacuum.  The transition
probabilities for a detector on a constant $U$-worldline are
determined from the Green functions in the global $(\tau, \sigma)$
vacuum $|0_{\rm Global} \rangle$.  It follows from \trans\ that with
respect to the proper time $\tau_D$ along the detector worldline
these are thermal Green functions, simply because the $(\tau,
\sigma)$ coordinates are invariant under imaginary shifts $t \to t +
{i \over T_H}$.  Accounting for the difference between $t$ and proper
time $\tau_D$, the detector sees a thermal bath of particles at
temperature $\sqrt{g^{00}}T_H = {1 \over 2 \pi Q} \sinh (2 \pi T_H
x)$.  The vacuum with this property is known as the Hartle-Hawking
vacuum.  Hence we conclude that
\eqn\hhg{
|0_{\rm Global} \rangle = |0_{\rm Hartle-Hawking} \rangle.
}

Yet another way to define a vacuum is as the state annihilated by
modes which are positive frequency in the Poincar\'e metric
\eqn\metp{
{ds^2 \over Q^2 L_p^2} = {-dT^2 + dy^2 \over y^2}.
}
We use capital $T$ to distinguish the Poincar\'e time $T$ from the
Schwarzschild time $t$.  For $-\infty < T < \infty$ and $0 < y <
\infty$ these coordinates cover only the patch defined by $\tau +
\sigma < {\pi \over 2}$ and $\tau - \sigma > - {\pi \over 2}$, and
hence only the boundary at $\sigma = - {\pi \over 2}$ (the various
coordinate systems are illustrated in figure 3).  These
coordinates are related to the global coordinates by the
transformation
\eqn\rlt{
T \pm y = \tan \half (\tau \pm \sigma \pm {\textstyle{\pi \over 2}}).
}
The (Klein-Gordon) overlap between a positive frequency mode in
Poincar\'e coordinates $\phi^P_{+ \omega} = {1 \over \sqrt{\pi
\omega}} e^{- i \omega T} \sin(\omega y)$ and a mode $\phi^G_{n} =
{1 \over \sqrt{\pi |n|}}
e^{- i n \tau} \sin(n(\sigma + {\textstyle{\pi
\over 2}}))$ with positive ($n = 1,2,\ldots$) or negative ($n = -1,
-2,\ldots$) frequency in global coordinates is
\eqn\wrto{
\vev{\phi^P_{+ \omega} | \phi^G_{n}}
= i \int_0^\infty dy \left[ \phi^P_{- \omega}
 (\p_T \phi^G_{n}) - \phi^G_{n} (\p_T \phi^P_{-\omega})\right]_{T=0},
}
where $\phi^P_{-\omega} = (\phi^P_{+\omega})^*$.
On the slice $T = 0$ one has $\sigma + {\pi \over 2} = 2 \tan^{-1} y$
and $\p_T = {2 \over y^2 + 1} \p_\tau$.
Using these facts and $\tan^{-1}y = {1 \over 2 i} \log
({1 + i y \over 1 - i y})$ one can put \wrto\ into the form
\eqn\wrtq{
\vev{\phi^P_{+ \omega} | \phi^G_{n}}
= {1 \over \pi} \sqrt{|n| \over \omega} \int_{-\infty}^\infty
dy\, e^{i \omega y}  (1 + i y)^{-n - 1}  (1 - i y)^{ n - 1}.
}
The contour must be closed in the upper half plane.  When $n$ is
negative there is no pole in the upper half plane and so the integral
vanishes.  When $n$ is positive there is a pole at $y = i$, and the
result of the integration is
\eqn\wrtr{\eqalign{
\vev{\phi^P_{+ \omega} | \phi^G_{+n}} &= (-1)^n \sqrt{n\over \omega}
e^{-\omega}
L_n^{-1} (2 \omega),\cr
\vev{\phi^P_{+ \omega} | \phi^G_{-n}} &= 0,
}}
where $L_n^\alpha$ is the associated Laguerre polynomial.
We conclude that the Bogoliubov transformation is block diagonal, and
it follows that the Poincar\'e annihilation operators are linear
combinations of the global annihilation operators and have no overlap
with the global creation operators, and hence
\eqn\hhg{
|0_{\rm Global} \rangle = |0_{\rm Poincare} \rangle.
}
This result will be confirmed by the computation of the Green
functions for massive scalars in section 6.
The equivalence of the global and Poincar\'e vacua in $AdS_n$ has
been discussed in \ulf.

\vskip .6cm
\centerline{\epsfxsize=0.7\hsize\epsfbox{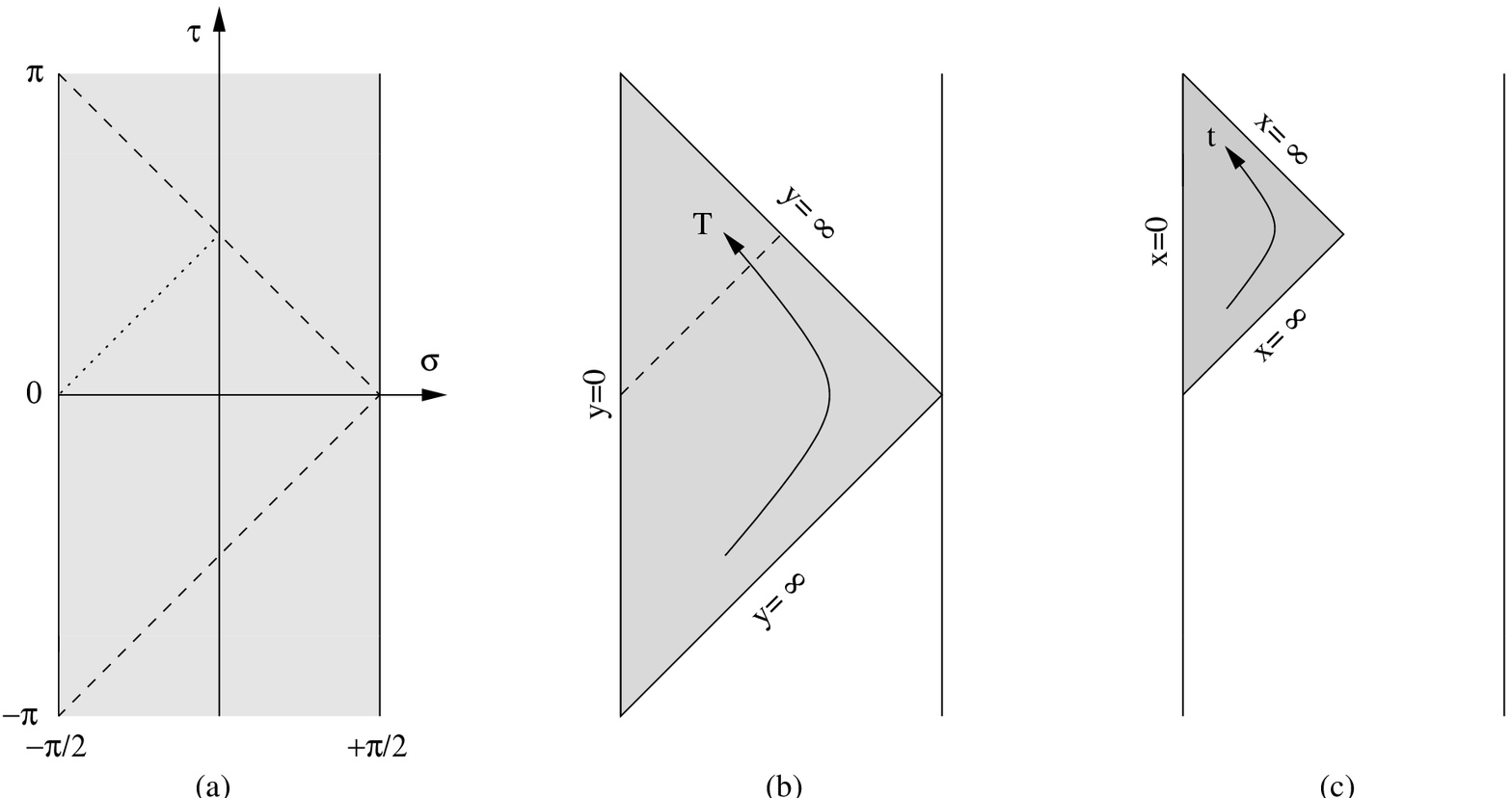}}
\bigskip
\centerline{\vbox{\noindent {\bf Fig.\ 3.}
Three coordinate systems on \ad.  (a) Global coordinates, $- {\pi
\over 2} \le \sigma \le {\pi \over 2}$ and  $-\infty <
\tau < \infty$.  (b)
Poincar\'e coordinates, $-\infty < T < \infty$, $0 < y < \infty$.
(c) Schwarzschild coordinates, $-\infty <  t < \infty$,  $0 < x <
\infty$.
}}
\vskip .7cm

We note that in the limit $T_H \to 0$, the Schwarzschild metric
\bgfe\ reduces to the Poincar\'e form
\eqn\fjjff{
{ds^2 \over Q^2 L_p^2} = {-dt^2 + dx^2 \over x^2}.
}
Hence the vacuum associated to the coordinates \bgfe\ reduces to the
$\slr$ invariant Poincar\'e vacuum associated to the coordinates
\fjjff\ in the limit $T_H \to 0$.  The Hawking temperature $T_H$ can
be thought of as a measure of the non-$\slr$ invariance of the vacuum
state associated to \bgfe.

\newsec{Entanglement Entropy}

The presence of a thermal bath of particles around an \ad\ black hole
would normally imply an associated temperature-dependent entropy.
However in the near-horizon limit \nhr, \fgh\ one finds that
\eqn\fot{
S_{BH} \to \pi Q^2,
}
independently of $T_H$.  This means that there is no classical
temperature-dependent entropy.  However at the one loop level there
is a quantum correction to the entropy from the entanglement of the
near-horizon \ad\ Hilbert space with the Hilbert space of the
decoupled asymptotically flat region.  (Strictly speaking
when this entropy
is nonzero the asymptotically flat region is not fully decoupled.)

In order to compute this entropy one needs to be more precise about
how the near-horizon \ad\ region of the Reissner-Nordstrom black hole
is separated from the asymptotically flat region in the $L_p \to 0$
limit.  Before taking $L_p$ all the way to zero let us choose a fixed
value of radial coordinate $U = U_{\rm max}$ which divides the
spacetime so that the \ad\ region is $0 < U < U_{\rm max}$ while the
flat region is $U_{\rm max} < U < \infty$.  $U_{\rm max}$ should be
in the mouth region where the geometry changes from \ad\ to flat, so
we take $U_{\rm max} = c_0 {Q \over L_p}$.  The arbitrary constant
$c_0$ can be taken to be very small so that the boundary is deep in
the \ad\ region, but is held fixed as $L_p \to 0$, so that $U_{\rm
max} \to \infty$.  We then erect the Hilbert space of, $e.g.$, a
scalar field on both regions, with bases denoted
$|\psi^i_{\rm AdS}\rangle$ and $|\psi^J_{\rm Flat}\rangle$.  A
generic state of the quantum field on the Reissner-Nordstrom
spacetime -- including the vacuum state -- is a sum of product states
of the form
\eqn\stte{
|\psi\rangle = \sum_{iJ} c_{iJ} |\psi^i_{\rm AdS}\rangle |\psi^J_{\rm
Flat}\rangle.
}
The state on the \ad\ region is then a density matrix
\eqn\dns{
\rho_{\rm AdS} = {\rm Tr}_{\rm Flat} |\psi \rangle \langle \psi | =
\sum_{ijK} c_{iK} c^*_{jK} |\psi^i_{\rm AdS}\rangle \langle
\psi^j_{\rm AdS}|.
}
Alternately the state on the flat region is
\eqn\dnns{
\rho_{\rm Flat} = \Tr_{\rm AdS} |\psi \rangle \langle \psi | =
\sum_{IJk} c_{kI} c^*_{kJ} |\psi^I_{\rm Flat}\rangle \langle
\psi^J_{\rm Flat}|.
}
The entanglement entropy is then defined by
\eqn\glu{
S_{\rm ent} = -\Tr \rho \ln \rho,
}
and takes the same value for either $\rho_{\rm Flat}$ or
$\rho_{\rm AdS}$.  $S_{\rm ent}$ is a measure of the correlation
between the portions of the quantum state on the two regions.

Entanglement entropy for black holes has been discussed in
\refs{\bls \mark \suss \cw \fiola-\kabat \larsen}.  In general there are
divergences arising from the entanglement of arbitrarily short
wavelength modes which overlap the dividing line $U_{\rm max}$.  We
are interested in finite, temperature-dependent contributions to
$S_{\rm ent}$ for the vacuum state on the Reissner-Nordstrom
geometry.  Such a term arises from the $S$-wave modes of scalar
fields, which reduces to a conformal field on \ad.  The vacuum
entanglement entropy for a conformal field theory of central charge
charge $c$ in curved space was derived in \refs{\fiola, \holzhey} as
\eqn\fla{
S_{\rm ent} = {c \over 6} \rho(\sigma_{\rm max}) - {c \over 6}
\ln \Delta.
}
In this expression, $\rho(\sigma_{\rm max})$ is the metric conformal
factor in the coordinate system used to define the vacuum evaluated
at the dividing line between the two regions, and $\Delta$ is a
non-universal short-distance cutoff.

The Hartle-Hawking vacuum for an \ad\ black hole is defined with
respect to the global coordinates \bgf, in which
\eqn\trop{
\rho = - \ln \cos \sigma.
}
For small $L_p$, $U_{\rm max}$ is large and from \ctf\ and \rgo\
we have
\eqn\sur{
\sigma_{\rm max} + {\pi \over 2} \sim {2 \pi Q^2 T_H \over
U_{\rm max}}.
}
It follows that
\eqn\rtl{
S_{\rm ent} = -{c \over 6} \ln (Q T_H) + {\rm non-universal}.
}
Related (although not obviously equivalent) results were obtained
with Euclidean methods in \ffgk.

Expression \rtl\ represents a logarithmic violation of decoupling in
the near horizon limit at finite temperature between the flat region
and the \ad\ region.  Additional contributions to the entanglement
entropy could arise from massive fields as well as higher angular
modes of massless fields.  However it is not clear if these
contributions will survive the near horizon limit since the modes of
such fields vanish rapidly near the boundary of \ad.  It would be
interesting to compute $S_{\rm ent}$ in string theory examples and to
investigate its origin in the D-brane picture.

\newsec{Making an \ad\ Black Hole}

In this section we consider simple processes which change the
temperature of the black hole.
A general spherically symmetric solution of Einstein-Maxwell gravity
corresponding to null matter falling in to a Reissner-Nordstrom black
hole is
\eqn\hoy{\eqalign{
ds^2 &= - {(r - r_+(v))(r - r_-(v)) \over r^2} dv^2 + 2 dr dv + r^2
d\Omega_2^2,\cr
F &= Q\epsilon_2,
}}
with $r_+ r_- = L_p^2 Q^2$.  The null matter has only one nonzero
component of its stress tensor:
\eqn\tvv{
T_{vv} = {\p_v r_+(v) + \p_v r_-(v) \over 4 \pi L_p^2 r^2}.
}
Let us start with an extreme Reissner-Nordstrom
black hole ($r_- = r_+$) and send in
a null shockwave of the form
\eqn\swv{
T_{vv} = {\pi Q^3 T_0^2 L_p \delta(v) \over r^2},
}
where $T_0$ is a constant with units of temperature.  The meaning of
this particular form will shortly be clear.  From \tvv\ we
see that this leads to
\eqn\leads{\eqalign{
r_+ + r_- &= 2 Q L_p ~~~~~ v < 0,\cr
r_+ + r_- &= 2 Q L_p + 4 \pi^2 Q^3 T_0^2 L_p^3 ~~~~~ v > 0.
}}
Using $r_+ r_- = Q^2 L_p^2$ one can solve to find
\eqn\isdf{\eqalign{
r_\pm &= Q L_p ~~~~~ v < 0,\cr
r_\pm &= Q L_p \left[ 1 \pm 2 \pi Q T_0 L_p \right] + O(L_p^3) ~~~~~
v > 0,
}}
where the higher-order corrections in $L_p$ will not be important.
We see then that the Hawking temperature $T_H ={ r_+ - r_- \over 4
\pi r_+^2}$ is
\eqn\hats{\eqalign{
T_H &= 0 ~~~~~ v < 0,\cr
T_H &= T_0 + O(L_p) ~~~~~ v > 0.
}}
The shockwave \swv\ increases the Hawking temperature of the black
hole from zero to $T_H$, at least in the $L_p \to 0$ limit.

Now we consider a near horizon limit
\eqn\ndhr{
L_p \to 0,
}
with
\eqn\fdgh{
U = {r - r_+ \over L_p^2},~~~~~~~Q,~T_0~~  {\rm fixed}.
}
The two-dimensional metric then reduces to
\eqn\dfrg{
{ds^2 \over L_p^2} = -{U(U + 4 \pi Q^2 T_0 \Theta(v)) \over Q^2} dv^2
+ 2 dU dv.
}
We note that in this limit the energy density \tvv\ vanishes.
In terms of the coordinates $s^\pm$ defined by
\eqn\erg{\eqalign{
s^- &= v,~~~~~~ s^+ = v + {2 \over U} ~~~~~ v < 0,\cr
s^- &= {1 \over 2 \pi Q^2 T_0}(e^{2 \pi T_0 v}-1), ~~~~~~ s^+ = s^- +
{2 \over U} e^{2 \pi T_0 v} ~~~~~ v > 0,
}}
the metric \dfrg\ takes the Poincar\'e form
\eqn\prf{
{ds^2 \over Q^2 L_p^2} = -{4 ds^+ ds^- \over (s^+ - s^-)^2}.
}
A detector at fixed $U = U_0$ hence has a worldline
\eqn\swv{\eqalign{
s^+ &= s^- + {2 \over U_0}, ~~~~~ s^- < 0,\cr
s^+ &= s^- (1 + {\textstyle{4 \pi Q^2 T_0 \over U_0}})
+ {2 \over U_0}, ~~~~~ s^- > 0.
}}
The proper time
$\tau_D$ along the detector worldline is
\eqn\taup{\eqalign{
\tau_D &= Q U_0 s^- ~~~~~ s^- < 0,\cr
\tau_D &= {1 \over 2 \pi Q T_0} \sqrt{U_0 (U_0 + 4 \pi Q^2 T_0)}
\ln (1 + 2 \pi Q^2 T_0 s^-) ~~~~~ s^- > 0.
}}
Since Poincar\'e time and worldline time are proportional prior to
the shock wave, there will be no particle detection in this region.
However, after the shock wave, it follows from \erg\ that $s^-$ is
periodic under imaginary shifts of detector proper time.  This
implies that the detector will detect a thermal bath of radiation at
temperature
\eqn\sto{
T = T_0 {Q \over \sqrt{U_0(U_0 + 4 \pi Q^2 T_0)}}.
}
The first factor of $T_0$
is the black hole temperature, while the second is
the Tolman factor representing the usual position-dependent
temperature for thermal equilibrium in a gravitational field.

In conclusion, \dfrg\ represents an \ad\ black hole whose temperature
grows as a function of the null coordinate $v$ because matter is
being thrown in.  A detector stationed at fixed $U$ outside the black
hole detects a thermal bath of radiation whose temperature grows as
the matter is thrown in.

\newsec{Massive Fields and Vacua}

In this section we extend the previous discussion to the case of
massive fields.  For the remainder of this paper we set $Q L_p = 1$.
The proper dependence may be restored using dimensional analysis.

\subsec{Green functions}

We consider a massive scalar field $\phi$ with action
\eqn\action{
S = -{1 \over 2} \int d^2 x \sqrt{-g} \left[ (\nabla \phi)^2 + m^2
\phi^2 \right].
}
The vacuum $|0\rangle$ is completely specified by the two-point
function $G({\bf x}, {\bf y}) = \vev{0|\phi({\bf x}) \phi({\bf y})
|0}$.  In Lorentzian spacetimes there are many Green
functions.\foot{A discussion can be found in chapter 4 of \full.}  We
focus on the Hadamard function
\eqn\had{
G^{(1)}({\bf x}, {\bf y}) = \vev{0 | \{ \phi({\bf x}),
\phi({\bf y})\}| 0},
}
which is related to the familiar Feynman propagator $G_F({\bf x},
{\bf y}) = i \vev{0 | T \phi({\bf x}) \phi({\bf y}) | 0}$ by $G^{(1)}
= 2\,{\rm Im}\,G_F$.  To construct the Hadamard function explicitly
for a given vacuum one first finds a complete set of positive
frequency solutions (i.e. $\phi_\omega \sim e^{-i \omega t}$ where
$t$ is the chosen time variable) of the massive wave equation
\eqn\wav{
\nabla^2 \phi_\omega = m^2 \phi_\omega,
}
normalized with respect to the Klein-Gordon inner product, which
in conformal gauge takes the form
\eqn\kgip{
\vev{\phi_\omega | \phi_\omega'} = i \int_\Sigma \left[ \phi_\omega^*
(\p_t \phi_{\omega'}) - (\p_t \phi_\omega^*) \phi_{\omega'} \right],
}
where the integral is taken over a constant-time slice $\Sigma$.  We
encounter bases $\{ \phi_\omega(y) \}$ defined on the half-plane
$y \ge 0$ which oscillate as $y \to \infty$ and hence are not
integrable.  These modes are normalized by requiring that
\eqn\nnn{
\phi_\omega(y) \to {1 \over \sqrt{\pi \omega}} \sin(\omega y +
\delta_\omega) ~~~~~ {\rm as}\ y \to \infty.
}
This gives the correct relativistic delta-function normalization
\eqn\rell{
\vev{\phi_\omega | \phi_{\omega'}} = 2 \omega \int_0^\infty dy\ 
\phi_\omega^* \phi_{\omega'} = \delta(\omega - \omega').
}
Once the modes are known, the Hadamard function is given by
\eqn\hadt{
G^{(1)}({\bf x}, {\bf y}) = 2\, {\rm Re} \int d\omega\, \phi_\omega^*
({\bf x}) \phi_\omega({\bf y}).
}
If the spectrum of $\omega$ is discrete then the integrals should be
replaced by sums.

\subsec{The global vacuum}

In this subsection we construct the Green function associated with
the global vacuum.  The wave equation for a massive scalar in global
coordinates is
\eqn\poss{
\left[ \cos^2\sigma\,(-\p_\tau^2 + \p_\sigma^2) - h(h-1) \right] \phi
= 0,
}
where we write $m^2 = h (h-1)$.  The normalized positive-frequency
solutions are \toshio
\eqn\ghj{
\phi_n = \Gamma(h) 2^{h-1} \sqrt{n! \over \pi \Gamma(n+2h)} e^{-i
(n+h) \tau} (\cos \sigma)^h C_n^h(\sin \sigma), ~~~~~ n = 0,1,\ldots,
}
where $C_n^h$ is the Gegenbauer polynomial (\gr\ {\bf 8.930}).  The
Hadamard function \hadt\ for the global vacuum is therefore
\eqn\glv{\eqalign{
&G^{(1)}_{\rm Global}(\tau_1, \sigma_1; \tau_2, \sigma_2) = {
\Gamma(h)^2
2^{2h-1} \over \pi} (\cos \sigma_1 \cos \sigma_2)^h\cr
&~~~~~~~~~~~~~~~~~~~~~~~~~~~\times \sum_{n=0}^\infty {n! \over
\Gamma(n+2h)} \cos \left[(n+h)
(\tau_1-\tau_2)\right] C_n^h(\sin \sigma_1) C_n^h(\sin \sigma_2).
}}
This sum appears in \pbm\ (the mode sum for $AdS_n$ is calculated
in \burgess) and gives
\eqn\dgh{
G_{\rm Global}^{(1)}(\tau_1, \sigma_1; \tau_2, \sigma_2) =
{\Gamma(h)^2 \over 2 \pi \Gamma(2h)} \, {\rm Re} \left[ (2/d_{\rm
Global})^h F(h,h;2h;-2/d_{\rm Global}) \right],
}
where
\eqn\gld{
d_{\rm Global}(\tau_1, \sigma_1; \tau_2, \sigma_2) = {\cos(\tau_1 -
\tau_2) - \cos(\sigma_1 - \sigma_2)\over \cos \sigma_1 \cos \sigma_2}
}
is the $\slr$ invariant distance function on \ad, in global
coordinates.  This is the known result \barf\ for the $\slr$
invariant Green function of a massive scalar on \ad.  This function
has the desired properties:  it satisfies the massive wave equation
\poss, has the correct short-distance singularity, $G_{\rm Global}^{
(1)} \sim -{1 \over \pi} \ln \left|\epsilon\right|$ for two points
separated by a distance $\epsilon$, and $G_{\rm Global}^{(1)} \sim
(\cos \sigma)^{h}$ as $\cos \sigma \to 0$.  For a massless scalar
$h = 1$ we recover
\eqn\mgg{
G^{(1)}_{\rm Global} =
{1 \over 2 \pi} \ln \left| 1 + {2 \over d_{\rm Global}}
\right| = -{1 \over 2
\pi} \ln \left| {\cos (\tau_1-\tau_2) - \cos (\sigma_1-\sigma_2)
\over \cos(\tau_1-\tau_2) + \cos(\sigma_1+\sigma_2)} \right|,
}
which is the correct massless Green function on the strip, as may be
seen by summing the massless Green function on the plane over a
collection of image field sources.  This is required by the conformal
invariance of a massless scalar.

One may explicitly check that the the same Green function is obtained
in Poincar\'e coordinates \metp, as expected from the equivalence of
the corresponding vacua.  The massive wave equation in Poincar\'e
coordinates for a positive-frequency mode $\phi = e^{-i \omega T}
\chi(y)$ is
\eqn\pop{
\left[ {\p^2 \over \p y^2} + \omega^2 - {h(h-1) \over y^2} \right]
\chi(y) = 0.
}
The normalized positive-frequency modes (which vanish at the boundary
$y = 0$) are
\eqn\pmoa{
\phi_\omega(T,y) = e^{-i\omega T}\sqrt{y\over 2} J_{h-1/2}(\omega y),
}
so that the Hadamard function for the Poincar\'e vacuum is
\eqn\hap{\eqalign{
G_{\rm Poincare}^{(1)}(T_1, y_1; T_2, y_2) &= \sqrt{y_1 y_2}
\int_0^\infty d \omega \cos \left[ \omega (T_1-T_2) \right] J_{h-1/2}
(\omega y_1) J_{h-1/2}(\omega y_2)\cr
&= {\Gamma(h)^2 \over 2 \pi \Gamma(2h)}\, {\rm Re} \left[ (2/d_{\rm
Poincare})^h F(h,h;2h;-2/d_{\rm Poincare})\right],
}}
(the integral appears in \pbm) where
\eqn\gldt{
d_{\rm Poincare}(T_1, y_1; T_2, y_2) = {-(T_1-T_2)^2 + (y_1 - y_2)^2
\over 2 y_1 y_2}
}
is the $\slr$ invariant distance function in Poincar\'e coordinates.
\hap\ agrees precisely with \dgh\ as anticipated.  For a massless
scalar ($h = 1$) we recover
\eqn\masp{
G_{\rm Poincare}^{(1)} = - {1 \over 2 \pi} \ln \left| {-(T_1-T_2)^2
+ (y_1-y_2)^2 \over -(T_1 - T_2)^2 + (y_1 + y_2)^2}\right|,
}
which is the usual massless Green function on the half plane, as
required by conformal invariance.  The term in the denominator can be
thought of as coming from an image field source at $y_2' = -y_2$.

\subsec{The Boulware vacuum}

In this subsection we construct the Boulware Green function.  For
convenience we temporarily set $2 \pi T_H = 1$.  One can restore
$T_H$ simply by taking $(t,x) \to 2 \pi T_H (t,x)$.  The massive wave
equation for a positive frequency solution $\phi_\omega = e^{-i
\omega t} \phi(x)$ reads
\eqn\rmw{
\left[ {\p^2 \over \p x^2} + \omega^2 - {h(h-1) \over \sinh^2 x}
\right] \phi_\omega(x) = 0.
}
The solution which vanishes at $x = 0$ is
\eqn\unno{
\phi_\omega(t,x) = \sqrt{\omega \over 2} {\Gamma(h + i \omega) \over
\Gamma(1 + i \omega)} e^{-i \omega t} (\sinh x)^{1/2} P^{\ha-h}_{
-\ha-i \omega}(\cosh x),
}
where $P$ is the associated Legendre function and we have normalized
according to \nnn.  This gives the Hadamard function
\eqn\harr{\eqalign{
&G^{(1)}_{\rm Boulware}(t_1, x_1; t_2, x_2) = \left(\sinh x_1 \sinh
x_2 \right)^{1/2}\cr
&\times
\int_0^\infty \omega d \omega \left| {\Gamma(h + i \omega) \over
\Gamma(1 + i \omega)}\right|^2 \cos\left[\omega(t_1-t_2)\right]
P^{\ha-h}_{-\ha-i \omega}(\cosh x_1) P^{\ha-h}_{-\ha+i \omega}
(\cosh x_2).
}}
This integral cannot be evaluated in terms of elementary functions.
For the massless case $h = 1$ we have
\eqn\leg{
(\sinh x)^{1/2} P^{-\ha}_{-\ha \pm i \omega}(\cosh x)
=\sqrt{2 \over \pi} {\sin \omega x \over \omega},
}
and hence
\eqn\leh{\eqalign{
G^{(1)}_{\rm Boulware}(t_1, x_1; t_2, x_2) &= {2 \over \pi}
\int_0^\infty {d
\omega \over \omega} \cos \left[\omega(t_1 - t_2) \right]
\sin \omega x_1 \sin \omega x_2\cr
&= - {1 \over 2 \pi} \ln \left| {-(t_1 - t_2)^2 + (x_1 - x_2)^2
\over - (t_1 - t_2)^2 + (x_1 + x_2)^2} \right|,
}}
which again is the correct massless Green function on the half plane.
Since it is impossible to rewrite \leh\ as a function of the $\slr$
invariant distance
\eqn\gdttt{
d_{\rm Boulware}(t_1, x_1; t_2, x_2) =
{-\cosh(2 \pi T_H (t_1-t_2)) + \cosh(2 \pi
T_H (x_1-x_2)) \over \sinh (2 \pi T_H x_1) \sinh (2 \pi T_H x_2)},
}
we discover that the Boulware vacuum is not $\slr$ invariant.  In
particular, it is distinct from the global vacuum.

Using a recursion relation satisfied by the Legendre functions one
can write down a (very complicated) expression which gives the value
of the integral \leh\ for any positive integer $h$ in terms of sums
of logarithms and exponential-integral functions $Ei(z)$ (\gr\ {\bf
8.211}).  The formulas involved are lengthy and not illuminating.
For example, for $h = 2$ one finds
\eqn\htwo{
G^{(1)}_{\rm Boulware} = (\coth x_1 \coth x_2) G^{(1)}_{B,(h=1)} -
{1 \over 4 \pi}
\sum_{a,b,c=\pm 1} {Ei(a(t_1-t_2) +b x_1 +c x_2) \over e^{a(t_1-t_2)}
\sinh b x_1 \sinh c x_2}.
}
One can check that $G^{(1)}_{\rm Boulware}$ constructed in
this way satisfies the massive wave equation \rmw, has the correct
short-distance singularity $G^{(1)}_{\rm Boulware} \sim - {1 \over
\pi} \ln \left| \epsilon \right|$, and vanishes as $x^h$ when $x \to
0$.  These properties ensure that the Boulware vacuum is a `good'
vacuum, although it is singular along the horizon at $x = \infty$.

Furthermore, by restoring $(t,x) \to 2 \pi T_H (t,x)$ one can verify
that in the limit $T_H \to 0$, the Hadamard function for the
Boulware vacuum reduces to that of the global vacuum \hap\ (with
$(T,r)$ replaced by $(t,x)$), in agreement with the fact that the
coordinate systems coincide for $T_H = 0$ \fjjff.
Thus the Hawking
temperature $T_H$ is a measure of the non-$\slr$ invariance of the
Boulware vacuum.

\newsec{The Stress Tensor}

The various vacua in \ad\ are characterized by differing stress
tensor expectation values.  In this section we compute these
expectation values for both the massless and the massive case.

\subsec{Two-dimensional Rindler and Minkowski space}

We begin with a review of some well-known features of the
thermodynamics of two-dimensional Rindler space.  This will clarify
the meaning of the various \ad\ expressions.  Readers familiar with
this topic should skip to the next subsection.

The Rindler metric
\eqn\rme{
ds^2 = - e^{\kappa (U^+-U^-)} dU^+ dU^-
}
is related to the Minkowski metric $ds^2 = -du^+ du^-$ by the
coordinate transformation
\eqn\trn{
U^\pm = \pm {1 \over \kappa} \ln (\pm \kappa u^\pm),
}
where $\kappa$ is a constant.  Lines of constant $U^+ - U^-$
correspond to the worldlines of observers undergoing constant proper
acceleration $\kappa$ (see figure 4).

Consider a massless scalar field in Minkowski space.  We may
construct the stress tensor operator $T_{\mu \nu}$ normal-ordered
with respect to Minkowski coordinates, $u^\pm$, or with
respect to Rindler coordinates $U^\pm$.  These two operators
are related by the well-known formula
\eqn\npp{
T_{++}(U^+) = \left( {\p u^+ \over \p U^+} \right)^2 T_{++}(u^+)
+ {1 \over 12 \pi} \sqrt{\p u^+ \over \p U^+} {\p^2 \over \p {U^+}^2}
\sqrt{\p U^+ \over \p u^+}.
}
Here and henceforth the stress tensor in a given coordinate system is
always normal-ordered with respect to that coordinate system.
The difference in the two stress tensors reflects the fact that
observers which are stationary with respect to different coordinate
systems detect different particle densities.  Plugging in \trn\ 
gives
\eqn\fgh{
T_{++}(U^+) = e^{2 \kappa U^+} T_{++}(u^+)+ {\kappa^2 \over 48 \pi}.
}
Taking the expectation value of \fgh\ in the Minkowski vacuum gives
\eqn\jff{
\vev{T_{++}(U^+)}_M = {\kappa^2 \over 48 \pi},
}
which is the stress-energy density of a thermal bath of particles at
temperature $T = {\kappa \over 2 \pi}$.\foot{This temperature
is related to
the fact that the coordinate transformation \trn\ is periodic in
imaginary Rindler time with periodicity $\beta = {2 \pi \over
\kappa}$, so that any Green function constructed in Rindler
coordinates would also be periodic in imaginary Rindler time and
would therefore correspond to a thermal Green function at temperature
$\beta^{-1}$.}  This may be interpreted as radiation coming from the
Rindler horizon.  On the other hand, taking the expectation value of
\fgh\ in the Rindler vacuum gives
\eqn\jaff{
\vev{T_{++}(u^+)}_R = - {1 \over 48 \pi (u^+)^2},
}
which can be viewed as a divergent
Casimir energy arising from the presence of a
boundary at the Rindler horizon $u^+=0$.

So far we have ignored the other independent component of
$\vev{T_{\mu \nu}}$, which is determined by the trace anomaly formula
\eqn\tracc{
\vev{T_{+-}} = {1 \over 2} g_{+-} \vev{T} = {R \over 48 \pi} g_{+-}.
}
This vanishes for Rindler/Minkowski space but plays a role in \ad,
where $R = -2$.

The stress tensor for massive scalars in Rindler
space has been constructed in
\mannste.

\vskip .6cm
\centerline{\epsfxsize=0.25\hsize\epsfbox{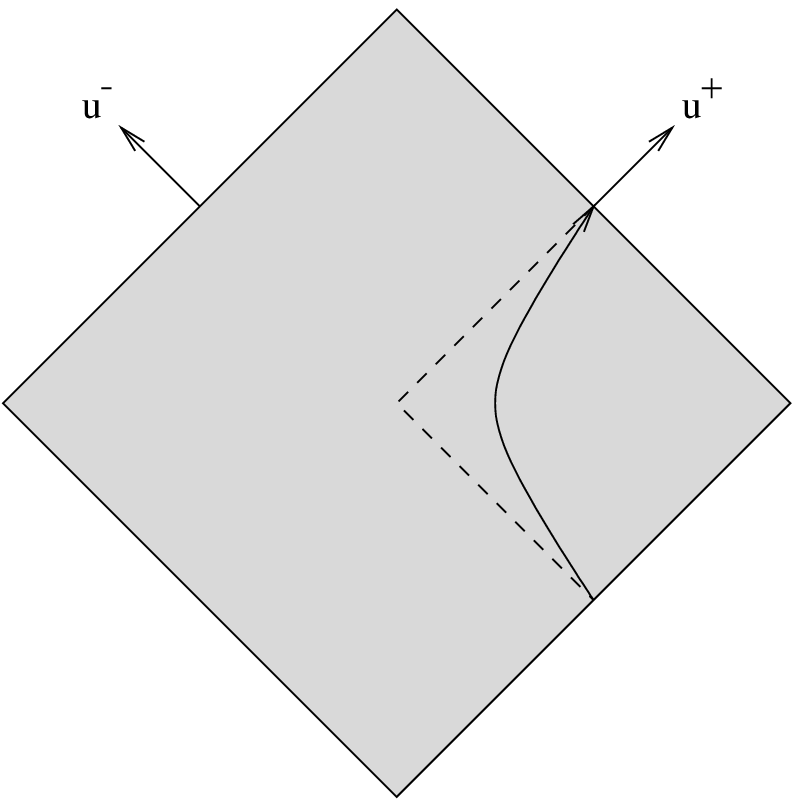}}
\bigskip
\centerline{\vbox{\noindent {\bf Fig.\ 4.}
Rindler spacetime.  The ``right Rindler wedge'' ($u^- < 0$ and
$u^+ > 0$) is accessible to a Rindler observer that accelerates
uniformly to the right.  The dashed lines show the past and future
horizon (the ``Rindler horizon'') seen by such an observer.
}}

\subsec{Massless scalar in \ad}

We now calculate the stress energy for a massless scalar in \ad.
The results are essentially identical to those we obtained in
the previous subsection.
It is convenient to work in null coordinates in which the
Poincar\'e and Schwarzschild coordinate systems take the form
\eqn\mmm{
ds^2 = -{4 du^+ du^- \over (u^+ - u^-)^2} = - {(2 \pi T_H)^2 dU^+
dU^- \over \sinh^2 \left[\pi T_H(U^+ - U^-) \right]},
}
where the null coordinates are defined by
\eqn\hhh{
2 \pi T_H U^\pm = 2 \pi T_H (t \pm x) = \ln(T \pm y) = \ln u^\pm.
}
 From the coordinate transformation \hhh\ we
find
\eqn\nop{
T_{++}(U^+) = (2 \pi T_H u^+)^2 T_{++}(u^+) +
{\pi T_H^2 \over 12},
}
where $T_{++}(U^+)$ is the stress tensor normal-ordered in
the Schwarzschild coordinates and $T_{++}(u^+)$ is the
stress tensor normal-ordered in the Poincar\'e coordinates.  Taking
the
expectation of this equation in the global vacuum gives
\eqn\ppp{
\vev{T_{++}(U^+)}_{\rm Global} = {\pi T_H^2 \over 12},
}
which is the stress-energy density of a thermal bath of particles at
temperature $T_H$ (again this is to be expected by virtue of the
periodicity of the coordinate transformation \hhh\ in imaginary
imaginary Schwarzschild time).  On the other hand, taking the
expectation value of \nop\ in the Boulware vacuum gives
\eqn\pssd{
\vev{T_{++}(u^+)}_{\rm Boulware}=-{1 \over 48 \pi (u^+)^2},
}
which may be viewed as Casimir energy arising from a boundary at the
black hole horizon.

So far we have discussed the stress tensor in Schwarzschild and
Poincar\'e coordinates. In null global coordinates
\eqn\glopl{
\half(\tau \pm \sigma \pm {\textstyle{\pi \over 2}})
=\tau^\pm=\tan^{-1}u^\pm,}
the stress tensor picks up a term
\eqn\stru{
T_{++}(\tau^+) = \left({\p u^+ \over \p \tau^+}\right)^2T_{++}(u^+) -
{1\over 12 \pi },}
so that
\eqn\prp{
\vev{T_{++}(\tau^+)}_{\rm Global} = -{1 \over 12 \pi },
}
which is the familiar zero-point  shift of a $c=1$
theory on the strip.
Curiously, normal-ordering in Poincar\'e and global coordinates lead
to different shifts even though the associated vacua are identical.
This is possible because in the former case one uses a continuous set
of modes, while in the latter one uses a discrete set, and so the
infinite zero-point energy sums are regulated differently.  The fact
that the expectation value of the stress tensor in the global vacuum
vanishes in Poincar\'e coordinates
but not in global coordinates also follows
from $\slr$ invariance, together with the observation that the
inhomogenous term in \npp\ vanishes for $\slr$ transformations
in Poincar\'e coordinates but not in global coordinates.

\subsec{Point-splitting regularization of massive scalars}

The calculation of $\vev{T_{\mu\nu}}$ for a massive scalar is
significantly more difficult as there is no simple formula such as
\npp.  The calculation is complicated by the fact that the
expectation value of an operator such as $T_{\mu \nu}$ which is
quadratic in the field $\phi$ is formally divergent and must be
regularized and renormalized. We implement the regularization by
using the point-splitting technique\foot{See \bd\ for a detailed
discussion.}, reviewed briefly below.

The stress tensor for a massive scalar field $\phi$ is
\eqn\tmunu{
T_{\mu \nu}({\bf x}) = \p_\mu \phi \p_\nu \phi - \ha
g_{\mu \nu} \left( g^{\rho \sigma} \p_\rho \phi \p_\sigma \phi + m^2
\phi^2 \right).
}
In conformal gauge
\eqn\metric{
ds^2 = -e^{2 \rho} dw^+ dw^-,
}
one has
\eqn\aaaa{\eqalign{
T_{++} &= \p_+ \phi \p_+ \phi, ~~~~~ T_{--} = \p_- \phi \p_- \phi,\cr
T_{+-} &= T_{-+} = - \ha g_{+-} m^2 \phi^2.
}}
Following \davies, we define the point-split stress tensor operator
as follows.  Consider any non-null geodesic through ${\bf x}$, and
let $x^\mu(\epsilon) = (w^+(\epsilon), w^-(\epsilon))$ be the point
on the geodesic at a proper distance $\epsilon>0$ from ${\bf x}$.
The geodesic may be characterized by its normalized tangent vector at
${\bf x}$, $\tau_0^\mu \equiv \tau^\mu(0)$, where
\eqn\aaab{
{d x^\mu(\epsilon) \over d \epsilon} = \tau^\mu(\epsilon), ~~~~
\tau_\mu \tau^\mu = -e^{2 \rho} \tau^+ \tau^- \equiv \Sigma = \pm 1.
}
The geodesic equations may be solved for $w^+$ in a power
series in $\epsilon$, giving
\eqn\reee{\eqalign{
w^+(\epsilon) &= w^+_0 + \epsilon \tau_0^+ - \epsilon^2 (\p_+ \rho)
(\tau_0^+)^2\cr
&~~~~~~~ + {1 \over 3} \epsilon^3 \left[(4
(\p_+ \rho)^2 - \p_+^2 \rho) (\tau_0^+)^3
- \p_-\p_+ \rho\tau_0^- (\tau_0^+)^2\right] + O(\epsilon^4),
}}
where $\rho$ on the right-hand side is always evaluated at
$w_0$.
Switching $+$ and $-$ in this expression yields the solution for
$w^-(\epsilon)$.
We define the point-split stress tensor operator by
\eqn\psto{\eqalign{
T_{++}({\bf x};\epsilon,\tau_0^\mu) &=U_\epsilon U_{-\epsilon}\ha
\left\{ \p_+ \phi({\bf x}(\epsilon)),
\p_+ \phi({\bf x}(-\epsilon)) \right\}\cr
T_{+-}({\bf x};\epsilon,\tau_0^\mu)
&= - \ha m^2 g_{+-}\left\{
\phi({\bf x}(\epsilon)), \phi({\bf x}(-\epsilon)) \right\},
}}
and similarly for $T_{--}$. In this expression
\eqn\dtu{
U_\epsilon \equiv \left({d w^+
(0) \over d\epsilon}\right)^{-1} {d w^+
(\epsilon) \over d\epsilon}.
}
These factors arises because $\p_+ \phi({\bf x}(\pm \epsilon))$
must be parallel transported back to ${\bf x}(0)$ in order to obtain
a quantity which transforms as a tensor \bd.
Upon taking the expectation value of both sides in some vacuum $V$,
we find that
\eqn\pstt{\eqalign{
\vev{T_{++}({\bf x};\epsilon,\tau_0^\mu)}_V &= \left[ U_\epsilon
U_{-\epsilon} {\p \over \p w_1^+} {\p \over \p w_2^+} \ha G_V^{(1)}
({\bf x}_1,{\bf x}_2)
\right]_{\scriptstyle {\bf x}_1 = {\bf x}(\epsilon)\ \ \, \atop
\scriptstyle {\bf x}_2 = {\bf x}(-\epsilon)},\cr
\vev{T_{+-}({\bf x};\epsilon,\tau_0^\mu)}_V &= \left[ -{m^2 \over 2}
g_{+-} {1 \over 2} G_V^{(1)}({\bf x}_1,{\bf x}_2)
\right]_{\scriptstyle {\bf x}_1 = {\bf x}(\epsilon)\ \ \, \atop
\scriptstyle {\bf x}_2 = {\bf x}(-\epsilon)}.
}}
and similarly for $\vev{T_{--}}$.

\subsec{Application of the point splitting procedure}

In all of the cases we consider the Hadamard function has the usual
short-distance behavior
\eqn\usual{
G^{(1)}(w^+_1,w^-_1;w^+_2,w^-_2) = -{1 \over 2 \pi} \ln \left|(w^+_1
-w^+_2)(w^-_1 - w^-_2) \right| + \cdots,
}
where the dots denote terms which are finite as ${\bf x}_2$
approaches ${\bf x}_1$, and the point-split stress tensors have the
general form
\eqn\ajb{\eqalign{
\vev{T_{++}({\bf x};\epsilon,\tau^\mu)} &= -{1 \over 4\pi}
\left[ {1\over \epsilon^2} -16 \Sigma\pi f_2({\bf x}) \right]
\tau_+ \tau_+ + f_1({\bf x})+O(\epsilon \ln \epsilon),\cr
\vev{T_{+-}({\bf x};\epsilon,\tau^\mu)} &= {m^2 \over 4 \pi} g_{+-}
\left[ \ln \epsilon  + f_3({\bf x})\right] +
O(\epsilon \ln \epsilon),
}}
where the three functions $f_1$, $f_2$, and $f_3$, which depend only
on the point ${\bf x}$ and not on $\epsilon$ or $\tau^\pm$, encode
all of the physical information in the point-split stress tensor.
To simplify the notation we here and henceforth drop
the subscript $0$ on $\tau^\mu$.
Finally, making use of the fact that $g_{++} = 0$ and
\eqn\flgg{
-\ha e^{2 \rho} = g_{+-} = 2 \Sigma \tau_+ \tau_-,
}
we can combine both expressions in \ajb\ into a single covariant
expression for the point-split stress tensor,
\eqn\onejy{\eqalign{
\vev{T_{\mu \nu}({\bf x};\epsilon;\tau^\mu)} &= {1 \over 8 \pi}
\left[{\Sigma \over \epsilon^2} -16 \pi f_2({\bf x}) \right]
(g_{\mu \nu}-2 \Sigma \tau_\mu \tau_\nu)+\theta_{\mu \nu}({\bf x})\cr
&~~~~~~~~~~+{m^2 \over 4 \pi} g_{\mu \nu} \left[ \ln \epsilon
+ f_3({\bf x}) \right] + O(\epsilon \ln \epsilon),
}}
where $\theta_{\mu \nu}$ is the traceless tensor whose components in
the $w^\pm$ coordinate system are
\eqn\compth{
\theta_{++} = \theta_{--} = f_1({\bf x}), ~~~~~
\theta_{+-} = \theta_{-+} = 0.
}

The regularized stress tensor $\vev{T_{\mu \nu}({\bf x};\epsilon,
\tau^\mu)}$ diverges in the limit $\epsilon \to 0$, and furthermore
the precise behavior of the divergence depends on the direction of
approach $\tau^\mu$.  The renormalized stress tensor is obtained
\davies\ by
discarding all of the terms in \onejy\ which depend explicitly on
either $\epsilon$ or $\tau^\mu$,
\eqn\reno{
\vev{T_{\mu \nu}({\bf x})} = g_{\mu \nu} \left[ {m^2 \over 4 \pi}
f_3({\bf x}) - 2 f_2({\bf x}) \right] + \theta_{\mu \nu}({\bf x}).
}
 From \onejy\ we see that the terms which diverge as $\epsilon \to 0$
are universal and do not depend upon the particular state under
investigation (i.e., they do not depend on the $f_i$).  Therefore
the divergent terms always cancel out when we calculate
the differences between stress tensors in different vacua.

\subsec{Energy of the global vacuum}

We begin by calculating $\vev{T_{\mu \nu}(u^+,u^-)}_{\rm Global}$ for the
$\slr$ invariant global vacuum in Poincar\'e coordinates.  The only rank
2 symmetric,
conserved, $\slr$ invariant tensor is $g_{\mu \nu}$, so we expect
that $\vev{T_{\mu \nu}}_{\rm Global} = c g_{\mu \nu}$ for some
constant $c$.  In the notation of the
previous subsection we find
\eqn\fff{
f_1 = 0, ~~~~~ f_2 = {1 + 3 h(h-1) \over 48\pi}, ~~~~~
f_3 = \psi(h)+\gamma.
}
where $\psi(z) = \p \ln \Gamma(z)/\p z$ and $\gamma = -\psi(1)$ is
Euler's constant.  Hence the renormalized stress tensor \reno\ is
\eqn\ppt{
\vev{T_{\mu\nu}}_{\rm Global} = {g_{\mu \nu} \over 2 \pi} \left[
-{1 \over 12}
- {h(h-1) \over 2} \left({1 \over 2} - \psi(h)-\gamma \right)\right].
}
We have obtained the same result by applying Pauli-Villars
regularization.  Note that when $h = 1$ we recover
\eqn\hffg{
\vev{T_{\mu \nu}} = -{g_{\mu \nu} \over 24 \pi},
}
which is the massless Weyl anomaly $\vev{T_{\mu \nu}} = {R \over 48
\pi} g_{\mu \nu}$, with $R = -2$ for \ad.

\subsec{Energy of the Boulware vacuum}

This calculation is significantly more complicated.  In particular,
we cannot use $\slr$-invariance to argue that $\vev{T_{\mu \nu}}_{
\rm Boulware}$ is proportional to $g_{\mu \nu}$, and indeed we find
that this is not the case.
To simplify the resulting expressions slightly we introduce
\eqn\ajk{
\vev{T_{\mu \nu}}' =
\vev{T_{\mu \nu}}_{\rm Global} - \vev{T_{\mu \nu}}_{
\rm Boulware},
}
with $\vev{T_{\mu \nu}}_{\rm Global}$ given by \ppt,
which is the energy difference between the global and Boulware vacua.
(Note that $\vev{T_{++}}' = -\vev{T_{++}}_{\rm Boulware}$ since
$\vev{T_{++}}_{\rm Global} = 0$.)
Using the Hadamard function
\harr\ constructed above, we find for $h=1,2,3$ the result
\eqn\expl{\eqalign{
\vev{T_{++}}'_{h=1} &= {\pi T_H^2 \over 12},\cr
\vev{T_{+-}}'_{h=1} &= 0,\cr
\vev{T_{++}}'_{h=2} &= {\pi T_H^2 \over 12}
\left[1 - 6\,{\rm csch}^2 z + 12 F(z) \,{\rm csch}^4 z \right],\cr
\vev{T_{+-}}'_{h=2} &= {g_{+-} \over 2 \pi}
\left[ 1 - \ln \left| {\sinh z \over z } \right| - 2 F(z)
\,{\rm csch}^2 z\right],\cr
\vev{T_{++}}'_{h=3} &= {\pi T_H^2 \over 12} \left[ 1 -
18\,{\rm csch}^2 z - 36 F(2 z)\,{\rm csch}^6 z +
18F(z)(3 \cosh 2 z + 5)\,{\rm csch}^6 z\right],\cr
\vev{T_{+-}}'_{h=3} &= {3 g_{+-} \over 2 \pi} \left[ {3 \over 2}
- \ln \left| {\sinh z\over z} \right| + {3 \over 2} F(2 z)\,
{\rm csch}^4 z - 6 F(z) \coth^2 z\,{\rm csch}^2 z\right],
}}
where we write $z = 2 \pi T_H x$ for simplicity.  We
have introduced the function
\eqn\fasxz{
F(w) = \int_0^w {du\over u} \sinh^2 u.
}

A conjectured expression for a general value of $h$ is
\eqn\rrv{\eqalign{
{\vev{T_{++}}' \over \pi T_H^2} &= {1 \over 12} - {h(h-1) \over 4
\sinh^2 z} \left[1 - h(h-1) \int_0^z {du \over u}
{\sinh^2 u \over \sinh^2 z} F\left(h+1,2-h,3,
{\sinh^2 u \over \sinh^2 z}\right)\right],\cr
\vev{T_{+-}}' &= {h(h-1)g_{+-} \over 4 \pi} \left\{ \psi(h)+\gamma
- \int_0^z du \left[ \coth u - {1 \over u}
F\left( h,1-h,1,{\sinh^2 u \over \sinh^2 z}\right)\right]\right\},
}}
where again $z = 2 \pi T_H x$.  Note that when $h$ is an integer the
hypergeometric series terminates, giving a polynomial which can be
explicitly integrated with relative ease (although the result is not
be expressible in terms of elementary functions but again involves
the exponential-integral function $Ei(z)$).  One can also check that
both components \rrv\ vanish as $T_H \to 0$, as should be expected.
Figures 5 and 6 show $\vev{T_{++}(x)}'$ and $\vev{T_{+-}(x)}'$ for
some values of $h$.

Evidence that \rrv\ is the correct expression for all values of $h$
is
\smallskip
{\narrower\smallskip\noindent
{a) Special cases.}
It correctly reduces to \expl\ for $h=1,2,3$.
\smallskip}
{\narrower\smallskip\noindent
{b) Conservation.}
The stress tensor should satisfy $\nabla^\mu T_{\mu \nu} = 0$, which
in Schwarzschild coordinates gives one nontrivial equation,\foot{The
other equation essentially says that $\vev{T_{\mu \nu}}$ should be
time-independent.}
\eqn\sfg{
{\pi^2 T_H^2 \over \sinh^2 (2 \pi T_H x)} {\p \over \p x} \vev{T}
+ {\p \over \p x} \vev{T_{++}} = 0,
}
where $\vev{T} = 2 g^{+-} \vev{T_{+-}}$,
which is indeed obeyed by \rrv.
\smallskip}

{\narrower\smallskip\noindent
{c) Behavior near the boundary.}
We saw earlier that the Schwarzschild modes behave like $\phi \sim
x^h$ near the boundary $x = 0$.  Therefore the stress tensor, which
is quadratic in $\p \phi$, should vanish as $\vev{T_{\mu \nu}} \sim
x^{2(h-1)}$ as $x \to 0$.  Again, this can be checked explicitly for
\rrv.  In particular, the physical requirement that $\vev{T_{\mu \nu
}}$ vanishes at the boundary fixes any overall additive constant,
and the fact that $g_{\mu \nu}$ diverges as $x^{-2}$ precludes us
from adding any constant multiple of the metric to $\vev{T_{\mu \nu
}}$.
\smallskip}

{\narrower\smallskip\noindent
{d) Behavior near the horizon.}
Finally we can consider the behavior near the horizon at $x \to
\infty$.  Everything becomes massless sufficiently close to the
horizon.  To see this, note that $g_{\mu \nu} \propto (\sinh x)^{-2}
\to 0$, in which case the Lagrangian density becomes
\eqn\acttt{
\CL = -{1 \over 2}  \sqrt{-g} \left[ (\nabla \phi)^2 + m^2 \phi^2
\right] \sim -{1 \over 2} \eta^{\mu \nu} \p_\mu \phi \p_\nu \phi,
}
where the inverse metric in the kinetic term $(\nabla \phi)^2$
cancels the zero coming from $\sqrt{-g}$.  Thus one should expect,
and we indeed find, that for $x \gg 1$ and any $h$ the expressions
expressions \rrv\ tend to the massless values
\eqn\massl{
\vev{T_{++}}' \to {\pi T_H^2 \over 12}, ~~~~~
\vev{T_{+-}}' \to 0, ~~~~~ x \gg 1.
}
This fixes the overall normalization of $\vev{T_{++}}$, which in turn
fixes the normalization of $\vev{T_{+-}}$ through the conservation
equation \sfg.
\smallskip}

\newsec{Boundary Correlation Functions}

It is expected \juan\ that string theory on \ad\ can be described as
conformally invariant quantum mechanics on the boundary of \ad.  The
conformal invariance of the $1$-dimensional boundary theory is a
consequence of the $SO(2,1)$ isometry group of \ad.  Boundary
correlation functions evaluated in any vacuum other than the natural
$SO(2,1)$ invariant vacuum, such as the Boulware vacuum, will
therefore not be conformally invariant.  However, we have seen that
the parameter $T_H$ is a measure of the non-$\slr$ invariance of the
Boulware vacuum, so we expect the nonconformal corrections to
boundary correlation functions in the Boulware vacuum to vanish as
$T_H \to 0$.  In this section we derive these boundary correlators
and verify that this is the case.

\subsec{Brief review}

In order to fix our conventions and notation we begin with a very
quick overview of the calculation of boundary correlation functions
using the bulk propagator.  The $AdS/CFT$ duality \juan\ states that
for every bulk field $\phi$ there is a corresponding local operator
$\CO$ on the boundary $\CB$, with
\eqn\aza{
Z_{\rm eff}(\phi) = e^{i S_{\rm eff}(\phi)} =
\vev{T e^{i \int_\CB \phi_{\rm b} \CO}},
}
where $S_{\rm eff}$ is the effective action in the bulk and $\phi_{
\rm b}$ is the field $\phi$ restricted to the boundary \refs{\gkp,
\witten}.  Let $\CO_h$ be the boundary operator of conformal weight
$h$ which couples to the bulk scalar $\phi$ of mass $m^2 = h(h-1)$,
and let $G_V(y,z;y',z')$ be the bulk two-point function of $\phi$ in
coordinates where the boundary lies at $y = 0$ and is parametrized by
$z$.  This could be Poincar\'e coordinates with $(y,z) = (y,T)$,
Schwarzschild coordinates with $(y,z) = (x,t)$, or global coordinates
with $(y,z)= (\cos \sigma, \tau)$.  The subscript $V$ is a reminder
that the two-point function $G_V$ expresses a choice of vacuum.
Boundary correlation functions will depend on the choice of
vacuum in the \ad\ bulk \refs{\albo, \albt}.

The two-point function should vanish as $y^h$ as either point
approaches the boundary, and we define the bulk-boundary propagator
for the corresponding vacuum state by \bdhm
\eqn\bbpl{
K_V(y,z;z') = \lim_{y' \to 0}\left[({y'})^{-h} G_V(y,z;y',z')\right].
}
(We ignore overall constants throughout this section).  If we are
given some boundary data $\phi_0(z')$ for the field $\phi$, then we
can use \bbpl\ to extend $\phi_0$ into the bulk by writing
\eqn\dfss{
\phi(y,z) = \int dz' K_V(y,z;z') \phi_0(z').
}
Then $\phi(y,z)$ satisfies the equation of motion in the bulk because
$K$ satisfies the equation of motion in the variables $(y,z)$.  Next
we plug the solution \dfss\ into the action \action.  Upon
integrating by parts, the action can be expressed as the boundary
term
\eqn\actt{
S = \lim_{ y\to 0} \left[
{1 \over 2} \int dz\  \phi(y,z) \p_y \phi(y,z) \right].
}
In the limit as we take the cutoff $y \to 0$ the bulk-boundary
propagator should approach a delta-function
\eqn\shodl{
K_V(y,z;z') \to y^{-h+1} \delta(z - z')
}
so we can replace
\eqn\repll{
\phi(y,z) \to y^{-h+1} \phi_0(z).
}
Then \actt\ becomes
\eqn\acdd{
S ={1 \over 2} \int dz\,dz'\,\phi_0(z) \phi_0(z')\left[ \lim_{y\to 0}
y^{-h+1} \p_y K_V(y,z;z')\right].
}
The generating function for correlation functions of $\CO_h(z)$ in
the boundary theory coupled to the source $\phi_0(z)$ is given by the
exponential of $i$ times \acdd, so recalling \bbpl\ we find that
(again, up to constants) \fmmr
\eqn\asddf{
\vev{\CO_h(z) \CO_h(z')}_V =  \lim_{y,y'
\to 0} \left[ ({y'})^{-h} y^{-h+1} \p_y  G_V(y,z;y',z') \right].
}

\subsec{Correlation functions in the global vacuum}

Substituting the global vacuum two-point function (in Poincar\'e
coordinates) \hap\ into \bbpl\ gives the familiar bulk-boundary
propagator
\eqn\pkkv{
K(y,T_1;T_2) = {y^h \over (y^2 - (T_1 - T_2)^2)^h},
}
which leads to the conformally invariant boundary correlation
function
\eqn\bpo{
\vev{\CO_h(T) \CO_h(0)}_{\rm Global} = {1 \over T^{2 h}}.
}

For purposes of comparison, it will be convenient to write \bpo\ in
Schwarzschild coordinates.  Recalling that the relation between
the Poincar\'e time $T$ and the Schwarzschild time $t$ on the
boundary is $2 \pi T_H t = \ln T$ and using the conformal
transformation law
\eqn\aazz{
\CO'(z') = (\p_z z')^{-h} \CO(z),
}
we can write \bpo\ in the form
\eqn\jdjj{
\vev{\CO_h(t) \CO_h(0)}_{\rm Global} = \left[ {T_H \over
\sinh{(\pi T_H t)} }\right]^{2h},
}
As expected, \bpo\ is
periodic in imaginary Schwarzschild time with periodicity $T_H^{-1}$
and therefore represents a thermal state at temperature $T_H$.  For
small separations \bpo\ has the universal UV limit $\vev{\CO_h(t)
\CO_h(0)} \sim {1 \over t^{2h}}$, while in the IR limit the
two-point function is exponentially suppressed due to the thermal
background, $\vev{\CO_h(t) \CO_h(0)}_{\rm Global} \sim e^{-2 \pi
T_H h t}$.

\subsec{Correlation functions in the Boulware vacuum}

Now we apply \asddf\ directly to the Boulware vacuum without first
constructing the Boulware bulk-boundary propagator $K_{\rm
Boulware}$ from \bbpl.  However, one can check that $K_{\rm
Boulware}$ is given by the Poincar\'e bulk-boundary propagator
\pkkv\ plus correction terms which are subleading in $z - z'$, so
that \shodl\ is still satisfied, and proportional to positive powers
of $T_H$, so that $K_{\rm Boulware}$ reduces to \pkkv\ as $T_H
\to 0$.

Using
\eqn\xxf{
(\sinh x)^{1/2} P^{\ha - h}_{-\ha \pm i \omega}(\cosh x) = {2^{\ha
-h} \over \Gamma(h + \ha)} x^{h} +  O(x^{h+2})
}
and the Boulware vacuum Green function \harr, we find from
\asddf\ that
\eqn\jww{
\vev{\CO_h(t) \CO_h(0)}_{\rm Boulware} =
\int_0^\infty \omega d\omega \left| {
\Gamma(h+i \omega) \over \Gamma(1+i \omega)}\right|^2 \cos \omega t,
}
where we have dropped all overall numerical constants.  This integral
is not convergent but may be defined by analytic continuation.  The
problem is that the limit \asddf\ does not commute with integration
over $\omega$.  We present a quick way of getting the answer, which
gives perfect agreement with a more careful analysis where one
computes the integral first and then takes the
limits.\foot{Alternatively, one may insert a factor of $e^{-\epsilon
\omega}$ into the integral \jww.  At least when $h$ is an integer,
the integral may be done explicitly, and the result is finite in the
limit $\epsilon \to 0$ and agrees precisely with the result we
present.}

Define $F_h(t)$ to be the quantity in \jww.  Then
\eqn\ddk{\eqalign{
F_{h+1}(t)
&= \int_0^\infty \omega d\omega \left| {\Gamma(h+1+i \omega)
\over \Gamma(1 + i \omega)}\right|^2 \cos \omega t\cr
&= \int_0^\infty \omega d\omega \left| {\Gamma(h+i \omega) \over
\Gamma(1 + i \omega)}\right|^2 (h^2 + \omega^2) \cos \omega t\cr
&= (h^2 - \p_t^2) F_{h}(t).
}}
This should be valid for all $h$.  To start the recursion we evaluate
\eqn\fjfjf{
F_1(t) = \left[ \int_0^\infty d \omega\ \omega^n \cos \omega t
\right]_{n=1}
= \left[- n!  t^{-n-1} \sin\left({n \pi \over 2}\right) \right]_{n=1}
= -{1 \over t^2},
}
where the quantity in brackets, which is strictly valid only for
$-1 < {\rm Re}(n) < 0$, is analytically continued to $n = 1$.  The
solution to \fjfjf\ and \ddk, up to ($h$-dependent!) constants, may
be summarized by the suggestive expression
\eqn\gig{
\vev{\CO_h(t) \CO_h(0) }_{\rm Boulware} = \left[{T_H \over
\sinh(\pi T_H t)}
\right]_{\rm singular}^{2h}
}
where we have restored the proper $T_H$-dependence.  The subscript
`singular' indicates that only the singular terms in the expansion of
the right-hand side of \gig\ around $t = 0$ are to be kept.  For
example, for $h = 3$ we find
\eqn\fssff{
\vev{ \CO_3(t) \CO_3(0) }_{\rm Boulware} \propto  {1 \over t^6}
- {\pi^2 T_H^2 \over t^4} + {8 \pi^4 T_H^4 \over 15 t^2}.
}

\vskip .7cm

\centerline{\bf Acknowledgements}

\vskip .2cm

It is a pleasure to thank V. Balasubramanian, R. Britto-Pacumio,
A. Chari, F. Larsen, A. Lawrence, Y-H. He, J. Michelson, I. Savonije,
J. Maldacena, S. Schmidt, and A. Volovich for many helpful
conversations.  This work was supported by an NSF
graduate fellowship and DOE grant DE-FG02-91ER40654.

\vskip .5in

\noindent
{\bf A. Appendix}

\vskip .6cm
\centerline{\epsfxsize=0.4\hsize\epsfbox{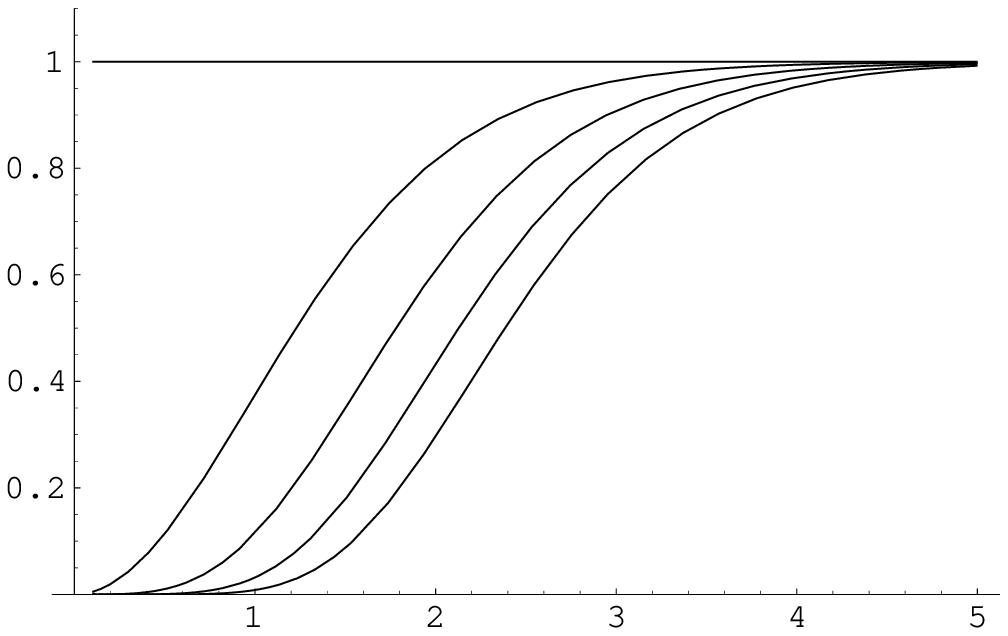}}
\bigskip
\centerline{\vbox{\noindent {\bf Fig.\ 5.}
Energy of a massive scalar in the Boulware vacuum.
This plot shows ${12 \over \pi T_H^2}
\vev{T_{++}(x)}'$, as defined in \ajk, as a function of $z = 2 \pi
T_H x$, for scalar fields of mass $h=1,2,3,4,5$ (from top to bottom).
}}

\vskip .5cm
\centerline{\epsfxsize=0.4\hsize\epsfbox{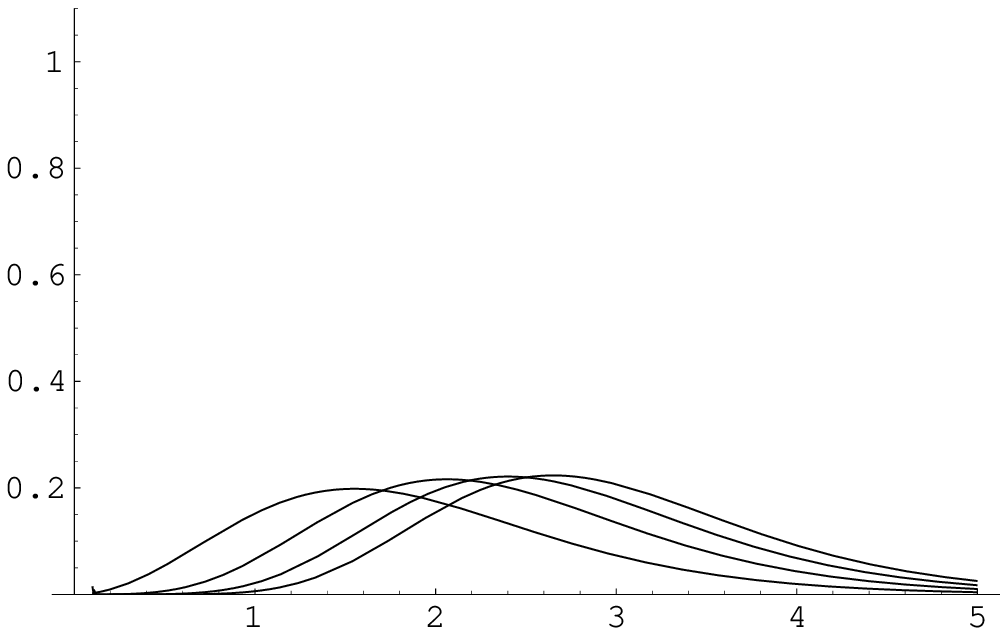}}
\bigskip
\centerline{\vbox{\noindent {\bf Fig.\ 6.}
This plot shows ${12 \over \pi T_H^2} \vev{T_{+-}(x)}'$
as a function of $z = 2 \pi T_H x$
for scalar
fields of mass $h=2,3,4,5$ (from left to right).
It vanishes identically for $h = 1$.
The scales are the same as in figure 5.
}}
\vskip .7cm

\listrefs

\end